\documentclass[review]{elsarticle}

\journal{Journal of Magnetism and Magnetic Materials}
\usepackage{float}
\usepackage{color}
\usepackage{amsmath}
\usepackage{mathrsfs} 
\usepackage{amssymb}
\usepackage{subfigure}
\usepackage{pslatex}
\usepackage{relsize}
\usepackage{bm}
\usepackage{xspace} 

\usepackage{xcolor}
\usepackage{soul}
\begin{document}
\begin{frontmatter}
\title{Analysis on the stability of in-surface magnetic configurations in toroidal nanoshells}
\author[Ufv]{A. W. Teixeira}
\author[UAC]{S. Castillo-Sep\'ulveda}
\author[PUC]{S. Vojkovic}
\author[Ufv]{J. M. Fonseca}
\author[Usach]{D. Altbir}
\author[Uchile]{\'A. S. N\'u\~nez}
\author[Ufv]{V. L. Carvalho-Santos}
\ead{vagson.santos@ufv.br}
\address[Ufv]{Universidade Federal de Vi\c cosa, Departamento de F\'isica, \\Avenida Peter Henry Rolfs s/n, 36570-000, Vi\c cosa, MG, Brasil}
\address[UAC]{Facultad de Ingenier\'ia, Universidad Aut\'onoma de Chile, Avda. Pedro de Valdivia 425, Providencia, Santiago, Chile, and CEDENNA, Avda. Ecuador 3493, Santiago, Chile}
\address[PUC]{Instituto de F\'isica, Pontif\'icia Universidad Cat\'olica de Chile, Campus San Joaqu\'in Av. Vicu\~na Mackena 4860, Santiago, Chile}
\address[Usach]{Departamento de F\'isica, Universidad de Santiago de Chile and CEDENNA,\\ Avda. Ecuador 3493, Santiago, Chile}
\address[Uchile]{Departamento de F\'isica, Facultad de Ciencias F\'isicas y Matem\'aticas, Universidad de Chile, Casilla 487-3, Santiago, Chile}

\begin{abstract}
Curvature of nanomagnets can be used to induce chiral textures in the magnetization field. Here we perform analytical calculations and micromagnetic simulations aiming to analyze the stability of in-surface magnetization configurations in toroidal nanomagnets. We have obtained that despite toroidal vortex-like configurations are highly stable in magnetic nanotori, the interplay between geometry and magnetic properties promotes the competition between effective interactions yielding the development of a core in a vortex state when the aspect ratio between internal and external radii of nanoturus is $\gtrsim0.75$.   
\end{abstract}

\begin{keyword}
{magnetic domains \sep nanotori \sep variable curvature \sep magnetization \sep vortex}
\end{keyword}

\end{frontmatter}


\def\ba{\begin{eqnarray}}
\def\ea{\end{eqnarray}}
\def\be{\begin{equation}}
\def\ee{\end{equation}}
\def\nn{\nonumber}
\bibliographystyle{apsrev}
\bibpunct{[}{]}{,}{n}{}{} 

\section{Introduction}

Magnetic configurations, such as vortices and skyrmions, can exhibit topological characteristics and the theoretical analysis of such magnetization patterns can be performed under the framework of the homotopy theory \cite{Nagaosa-Review}. Therefore, the statical and dynamical properties of these magnetic topological systems can present a strict relation with the geometry of the structure. In this frame, a research topic that is in fast development in nanomagnetism is the study of the magnetization properties of curved magnetic systems \cite{Gaididei-Review}. On this issue, it has been shown that the energy and stability of magnetic vortices and skyrmions depend on the geometry of the nanoparticle \cite{Kravchuk-PRB-2012,CSantos-JAP1,Vagson-JAP-2015}. In addition, curvature-induced effects in the energy, magnetic groundstate and excitations of magnetic nanoparticles have been analyzed in several works describing the  magnetization properties in curved nanomagnets as the sphere \cite{Goll-PRB-2004,New-sphere}, catenoid and hyperboloid \cite{CSantos-PLA-2013, CSantos-PLA-2012, CSantos-PLA-2013-2} and cones \cite{Freitas-PLA-2005}. 

A particular and  interesting geometry that appears to be a very suitable manifold is the torus. In fact, toroidal geometry is very adequate to study curvature effects in magnetic systems since it encompasses a interpolation between negative (pseudosphere or hyperbolic plane) and positive (sphere) Gaussian curvatures. This geometry has recently received a considerable attention in the literature and has been theoretically studied in different contexts. Some examples are  toroidal tight traps for Bose-Einstein condensation \cite{Bose-Einstein}, carbon nanotubes providing quasi-zero-dimensional systems whenever the rings are very small \cite{nanotubo1, nanotubo2}, and torus-shaped field-effect transistors for technological applications \cite{aplic-tecno}. From the theoretical point of view, toroidal geometry has been considered in the analysis of transport properties of topological insulators \cite{Jakson-EPJB-2016}, to determine dispersion relations and field patterns of the localized surface plasmons \cite{Mary-PRB-2007, Mary-PRB-2005}, to describe 2D-topological spin excitations \cite{C-Santos08}, and to perform molecular dynamics simulations of stable cyclotron motions of ions and water molecules in carbon structures \cite{Hazan-Mol-Torus}.   

Particularly in nanomagnetism, it has been shown that ferromagnetic nanotori can support a vortex configuration as the magnetization groundstate for smaller sizes than cylindrical nanorings \cite{CSantos-JAP1, Smiljan-JAP1}. The previous analytical works have considered a purely in-surface vortex configuration in a toroidal nanoparticle \cite{CSantos-JAP1}, which leads to a null dipolar energy associated to the vortex state. Nevertheless, micromagnetic simulations have revealed that small deviations from the azimuthal direction must produce magnetostatic charges and then, the vortex presents a nonnull dipolar energy \cite{Smiljan-JAP1}. The deviation from the purely in-surface state is associated with effective anisotropy and Dzyaloshinskii-Moriya interactions (DMI) induced by curvature \cite{Gaididei-PRL}. In fact, it has been shown that due to the variable curvature of the torus, the curvature-induced DMI generates  two  vortex-antivortex pairs as remanent state in hollow toroidal nanoparticles with high internal radius \cite{Smiljan-JAP2}.

In this work we study the magnetization on a rigid torus shell considering the effect of the curvature on the magnetic energy. We analyze static solutions considering the dipolar interaction as an on-surface anisotropy. More specifically, we look for general static solutions and show that the energy of a purely tangential magnetization configuration is minimized for an azimuthal vortex, while it is  maximized for a so called tokamak configuration. Indeed, the analytical solutions of the adopted model predicts a high stability of an in-surface azimuthal vortex, however, micromagnetic simulations have shown the possibility of development of an out-of-surface component of the magnetization. Thus, starting from an ansatz describing a vortex core model, we have shown that the energy of magnetic vortices can decrease by the development of a vortex core in a magnetic nanotori, as it occurs in their cylindrical counterparts \cite{Kravchuck-ring}. The region in which the vortex core appears depends on the geometrical parameters of the torus. From analytical calculations and micromagnetic simulations, we have then obtained a phase diagram showing regions of stability of purely in-surface state and a vortex with core.

This work is organized as follows: in section II we present the theoretical model to describe the magnetic states  and obtain the total energy functional in toroidal coordinates. Results are presented in section III where we get static solutions that describes a vortex and we discussed small fluctuations above this configurations to elucidate the role played by geometry in this topological excitations. In this section we also present micromagnetic results to show the development of an out-of-surface magnetization components and propose an ansatz to describe a vortex core model. Finally, in section IV the conclusions and prospects are presented. For more clarity, technical details are presented in the Appendix.

\section{Theoretical model}

\begin{figure}\begin{center}
\,\,\includegraphics[scale=0.4]{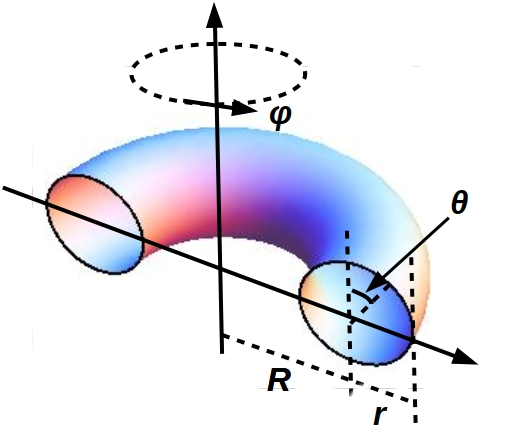}
\caption{Toroidal geometry and parameters describing a torus shell. 
$R$ and $r$ are the external and internal radii respectively and $\varphi$ and $\theta$ ($\in [0,2\pi)$)
are the azimuthal and polar angles,  respectively.}\label{Torus-Coordinates}\end{center}
\end{figure}

Our focus is the study of the properties of a magnetic toroidal shell. The simplest genus-1 torus is a ring-shaped geometry resembling a donut. Its geometry is described using a spherical-type coordinate system, with coordinates

\ba\label{CoordinateSystem}
x=\mathcal{Z}\cos\varphi\,,\,\,\,\,\,\,y=\mathcal{Z}\sin\varphi\,,\,\,\,\,\,z=r\cos\theta\,,
\ea
where $\mathcal{Z}=R+r\sin\theta$, $R$ and $r$ are  the external and internal radii,  respectively, and $\varphi$ and $\theta$ define the azimuthal and the polar-like angles, as shown in  Fig. \ref{Torus-Coordinates}.  The above set of equations allows to describe three different types of torus: the ring torus ($R>r$), the horn torus ($R=r$), and the self intersecting spindle torus ($R<r$). In addition, one can note that by taking $R=0$, the coordinate system describing the geometry of a sphere of radius $r$ is obtained.

In the framework of a continuum micromagnetic model, the magnetization profile can be written as a continuous function in a local basis, ($\hat{r},\hat{\theta},\hat{\varphi}$), 
in the form $\mathbf{M}=M_s\mathbf{m}(\theta,\varphi)\equiv M_s\mathbf{m}$, where $M_s$ 
is the saturation magnetization and
\ba\label{MagnProfile}
{\bf m}=\hat{r}\cos\Theta+\hat{\theta}\sin\Theta\cos\Phi+\hat{\varphi}\sin\Theta
\sin\Phi\,.
\ea
Here, $\Theta\equiv\Theta(\theta,\varphi)$ and $\Phi\equiv\Phi(\theta,\varphi,)$ 
describe the spherical polar and azimuthal angles of the magnetization vector field in a local base. 

In this work we will consider that the magnetic energy of any structure is given by $E_m=\mathcal{E}_{ex}+\mathcal{E}_{a}$, where $\mathcal{E}_{ex}$ and $\mathcal{E}_{a}$ are  the exchange and on-surface anisotropy contributions, respectively. The on-surface anisotropy can be viewed as a simplification on the evaluation of the magnetostatic energy based on the fact that in the regime where the thickness of the tube is small, the volumetric magnetostatic energy can be neglected and the surface magnetostatic energy can be captured by the inclusion of an \textit{ad hoc} superficial anisotropy \cite{Gaididei-PRL,Landeros-JAP}. The magnitude of such energy penalty can be estimated by the energy associated with a capacitor with a charge density proportional to the local magnetic charge ($\sigma=\mathbf{M}\cdot\mathbf{n}$). The global energy can be evaluated by accumulating the contribution of infinitesimal capacitors neglecting their interactions. On the other hand the exchange energy can be obtained from the model developed in a recent work of Gaididei \textit{et al.} \cite{Gaididei-PRL}, in which the authors show that when the magnetization vector is parametrized in the basis of a curved thin magnetic shell, the exchange energy density of an arbitrary surface can be written as 
\ba\label{EnFunc}
\mathcal{E}_{ex}=(\nabla\mathbf{m})^2=\left[\sin\Theta(\nabla\Phi -{\Omega})-
\cos\Theta\frac{\partial\Gamma(\Phi)}{\partial\Phi}\right]^2 +\left[\nabla\Theta-\Gamma(\Phi)\right]^2
\ea
where ${\Gamma}(\Phi)$ is a matrix depending on the Gauss and mean curvatures of the 
nanomagnet and ${\Omega}$ is a modified spin connection \cite{Gaididei-PRL,Gaididei-JPA}. 
By using the parametrization given in Eq. (\ref{CoordinateSystem}) and the definitions to 
${\Gamma}(\Phi)$ and ${\Omega}$ given in Ref. \cite{Gaididei-PRL} (see appendix \ref{Appendix1} for more details), we have that 
\be
\Gamma(\Phi)\equiv{\Gamma}=-\frac{\hat\theta}{r}\cos\Phi-\hat\varphi\frac{\sin\theta}{\mathcal{Z}}\sin\Phi\,,
\ee
and
\be
{\Omega} = -\frac{\cos\theta}{\mathcal{Z}}\hat{\varphi}\,.
\ee

Based on the above, the total magnetic energy functional of a ferromagnetic nanoshell is well captured by
\ba\label{EnTot}
E[\mathbf{M}]=h\int_S\left[\ell^2\mathcal{E}_{ex}+\lambda (\mathbf{M}\cdot\mathbf{n})^2\right]\,dS\,,
\ea
where $\ell\equiv\sqrt{A/(4\pi M_s^2)}$ is the exchange length, $h$ is the thickness of the 
magnetic shell, small enough to ensure magnetizations uniformity along its thickness, i. e. $\mathbf{m}=\mathbf{m}(\theta,\varphi)$, $dS$ is the surface element, 
$A$ is the exchange stiffness, and $\lambda$ is a normalized anisotropy coefficient.

\section{Results and discussion}

\begin{figure}
\includegraphics[scale=0.15]{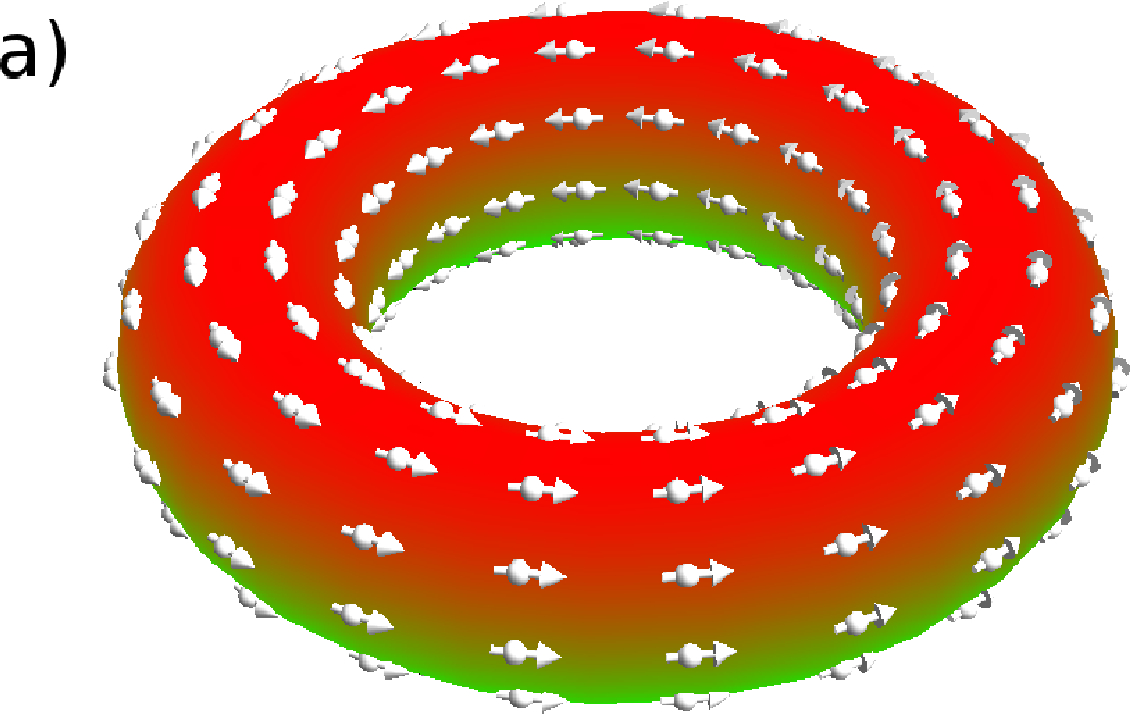}\includegraphics[scale=0.15]{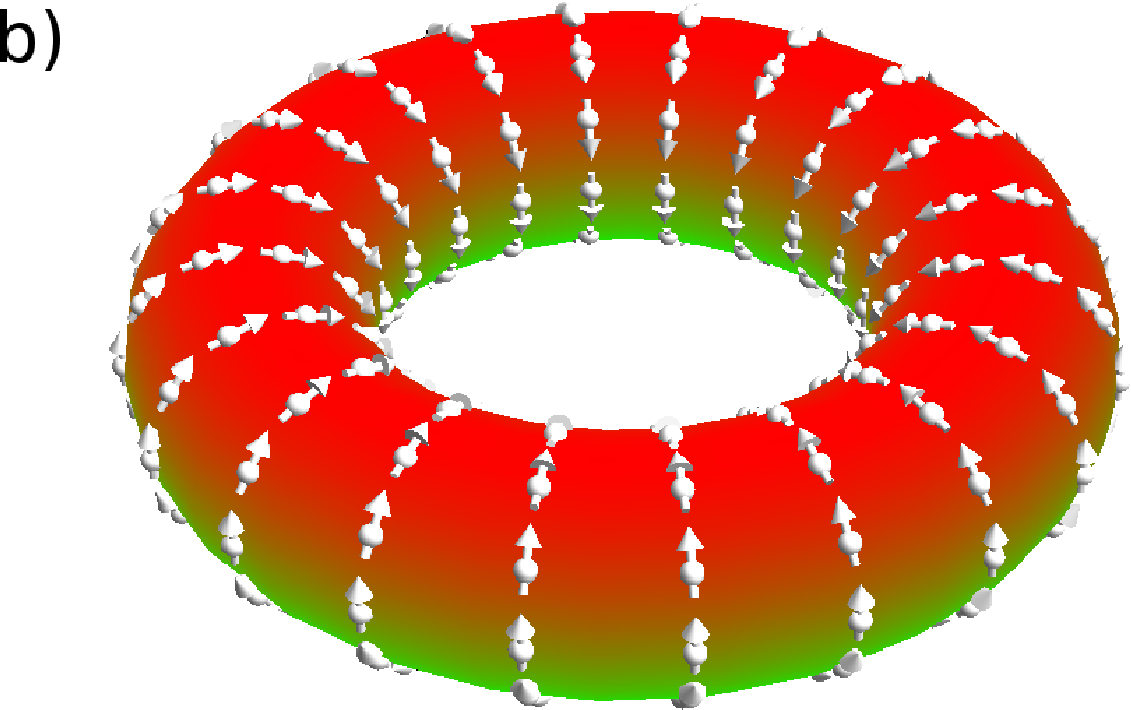}\includegraphics[scale=0.15]
{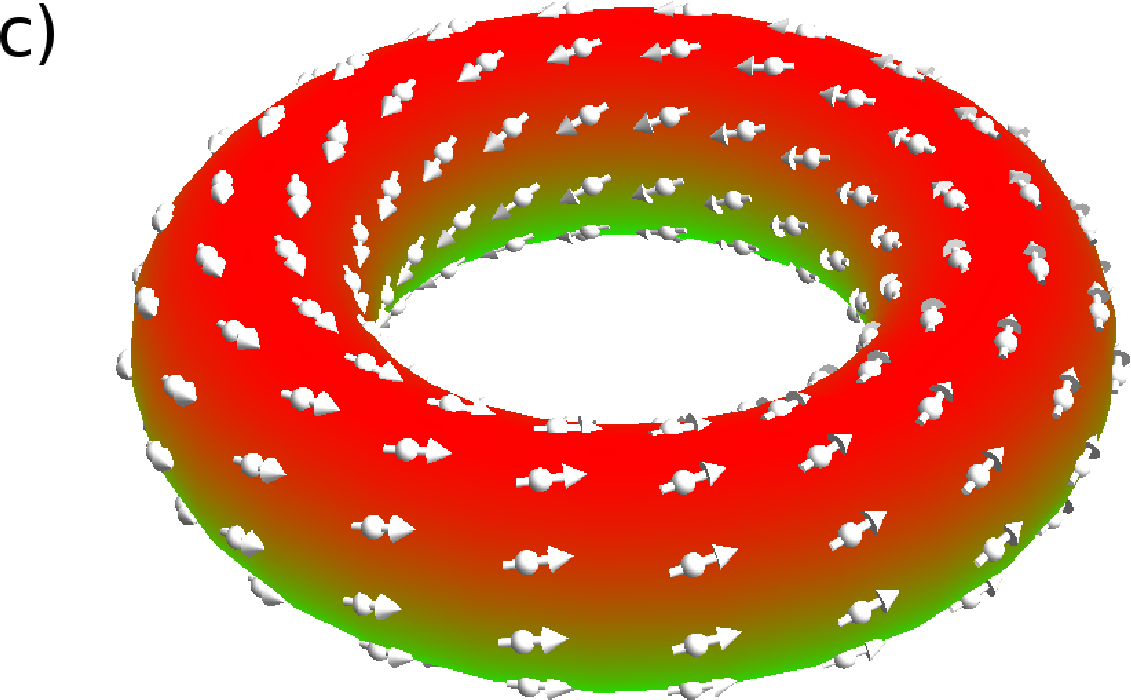}\caption{Three configurations representing in-surface states. 
Figures a, b and c represent respectively a toroidal vortex ($\alpha=\pi/2$), 
a poloidal vortex ($\alpha=0$) and a tokamak-like state with $\alpha=\pi/4$.}\label{ISConfigurations}
\end{figure}

Using expression (\ref{EnFunc}) for the exchange energy, we will analyze general static solutions for the case of an on-surface anisotropy ($\lambda>0$) with an anisotropy strong enough to provide a quasitangential magnetization distribution. However, some considerations about a purely on-surface configuration must be done. A tangential magnetization configuration can be obtained by taking $\Theta=\pi/2$ and $\Phi=\alpha$, where $\alpha$ is a constant. The magnetization profiles for an arbitrary $\alpha$ will be called tokamak-like (TK) configuration while magnetization configurations for $\alpha=\pi/2$ and  $\alpha=0$ will be called toroidal (TV) and poloidal vortex (PV) respectively. In  Fig. \ref{ISConfigurations} we represent the configuration profiles of TV, PV and TK for $\alpha=\pi/4$. A purely tangential magnetization configuration is then written as $\mathbf{m}=\hat{\theta}\cos\alpha+\hat{\varphi}\sin\alpha$ and the magnetic energy density given in Eq. (\ref{EnFunc}) is reduced to $\mathcal{E}^{t}=\Gamma^2+(\nabla\Phi-\Omega)^2$, and it is evaluated as 
\ba\label{TKDensity}
\mathcal{E}_{ex}=\frac{\cos^2\alpha}{r^2}+\frac{\sin^2\alpha\sin^2\theta}{\mathcal{Z}^2}+\frac{\cos^2\theta}{\mathcal{Z}^2}\,.
\ea
Therefore, the exchange energy of an in-surface state is given by 
\ba
{E}_{ex}={4\pi^2 \ell^2\,L}\Bigg[\frac{p}{1+\sqrt{1-p^2}}+\frac{p}{1-p^2+\sqrt{1-p^2}}\sin^2\alpha+\frac{1}{p}\cos^2\alpha\Bigg],
\ea
where $p=r/R$ is a variable that allows  to obtain a phase diagram  showing the stability of the in-surface states in function of $\alpha$. Our results for $p$ are shown in Fig. 3.a, evidencing that the exchage energy of the TK state is higher than TV or PV state for any $p$. In fact, one can note that for $p\lesssim 0.87$, the exchange energy is minimized for a TV state. However, when $p\approx0.87$, any value of $\alpha$, and consequently, any orientation of the magnetization of the on-surface state minimizes the exchange energy. Nevertheless, TK states can minimize exchange energy only in this specific aspect ratio of a nanotorus. The exchange energy of PV state starts to be lower than in the TV state even when $p\gtrsim0.87$. 

\begin{figure}\begin{center}
\includegraphics[scale=0.35]{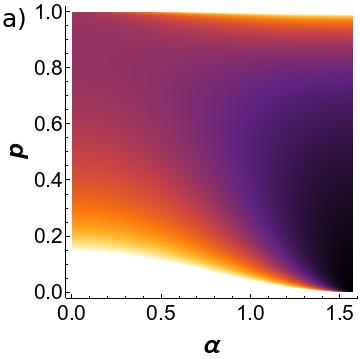}\hspace{0.5cm}\includegraphics[scale=0.4]{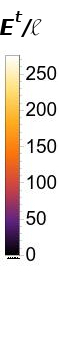}\\\includegraphics[scale=0.2]{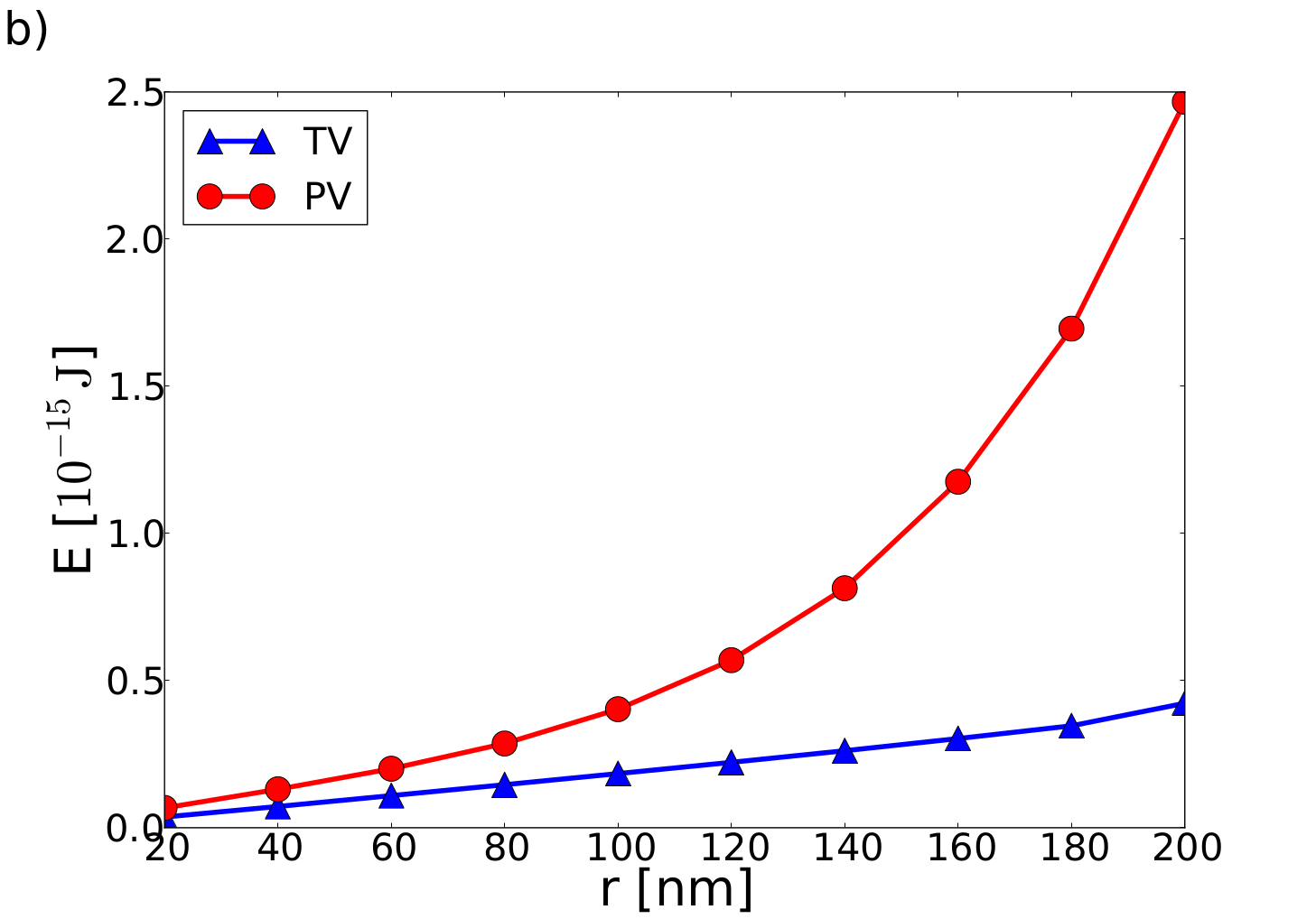}\caption{Fig. a shows a density plot of exchange energy of an in-surface state in function of $\alpha$ and $p$. Dark colors represents lower values of energy. Fig. b presents the total energy of a TV (blue) and a PV (red) states for $R=200$ nm in function of $r$.}\label{ToPoStateDiagram}\end{center}
\end{figure}

However, the minimization of the exchange energy does not ensure minimization of the total energy. Indeed, unlike TV configuration and despite it has not surface magnetic charges, PV state presents a dipolar contribution to the total energy coming from volumetric magnetic charges, formally defined as $\sigma=\nabla\cdot\mathbf{m}$. Therefore, this contribution must be taken into account in order to determine the in-surface magnetization profile that minimizes the total energy. In this context, we have performed micromagnetic calculations to determine the total energy of each state in permalloy tori. These calculations consider explicitly the magnetostatic interaction instead of an in-surface anisotropy in such way that the total energy is given by the sum of exchange and dipolar energies. For permalloy we used typical values of $A=13$ pJ/m, $M_s=0.84$ MA/m, and a damping constant of $0.5$.

Our results are summarized in Fig. \ref{ToPoStateDiagram}-b. It can be noticed that despite the PV configuration exhibit the lower  exchange energy for $r/R\gtrsim0.87$, it presents higher total energy due to the dipolar contribution. In this context, if we consider purely on-surface state, TV state is the lower energy configuration for any aspect ratio in a toroidal nanoparticle. 
    
\subsection{Vortex core}
The above presented solutions do not consider that there are components of the magnetization pointing along the out-of-surface direction. Then, aiming to determine possible deviation of the purely on-surface state  in magnetic nanotori, we make use of a continuous theory in which the magnetization density can be  written as
\begin{eqnarray}\label{MagProfile}
\mathbf{M}(\mathbf{r})\equiv\mathbf{M}=M_z(\theta,\varphi)\hat{z}+M_\varphi(\theta,\varphi)\hat{\varphi},
\end{eqnarray}
where $M_z^2+M_\phi^2=M_S^2$, with $M_S$ being the saturation magnetization.

A possible deviation of the in-surface state can be obtained in analogy with planar \cite{Landeros-dot} or curved dots \cite{Vagson-JAP-2015}, in which an out-of-plane component of the vortex state appears in the form of a core (VC). This configuration can be well described by $\mathbf{m}= \Omega_z \hat{z} + \Omega_\phi \hat{\phi}$, where $\Omega_\phi=\sqrt{1-\Omega_z^2}$. The profile $\Omega_z=[1-(\rho/\rho_c)^2]^4$ describes the vortex core, where $\rho=\mathcal{Z}$ and $\rho_c=R+r\sin\theta_c$. The angular parameter $\theta_c$ describes a critical poloidal angle in which magnetic moments start  pointing along the out-of-surface direction.  

\begin{figure}\begin{center}
\includegraphics[scale=0.28]{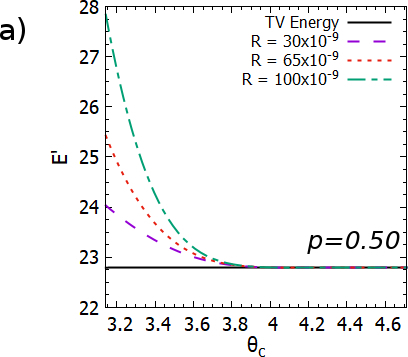}\hspace{0.2cm}\includegraphics[scale=0.28]{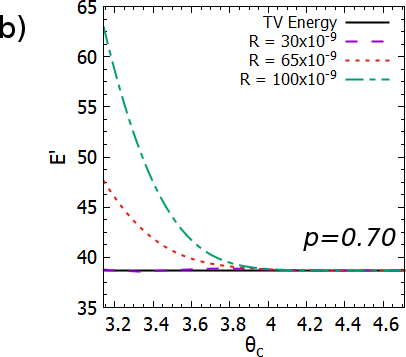}\\\includegraphics[scale=0.28]{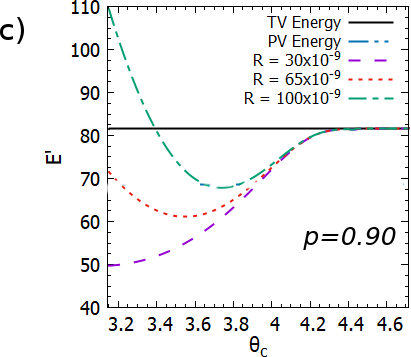}\hspace{0.2cm}\includegraphics[scale=0.28]{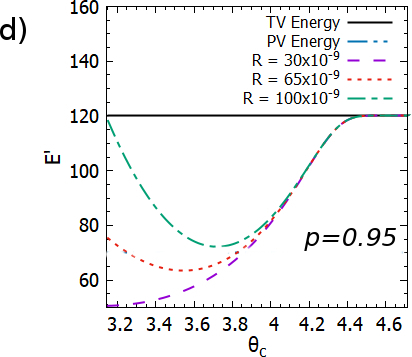}\caption{Magnetic energy of the considered magnetic states. Here, we present the energies for $r/R=0.50$ (a), $r/R=0.70$ (b), $r/R=0.90$ (c) and $r/R=0.95$ (d).}\label{EnergiesDifStates}\end{center}
\end{figure}

Assuming the magnetization field described by Eq. (\ref{MagProfile}), the exchange energy is given by
\begin{eqnarray}\label{ExchangeEqReduced}
E_{ex}= L\ell^2\int \left[\frac{1}{r^2(1-\Omega_z^2)}\left(\frac{\partial \Omega_z}{\partial\theta}\right)^2+\frac{1-\Omega_z^2}{\mathcal{Z}^2}\right]dS\,.\,\,\,\,\,\,\,\,
\end{eqnarray} 
The anisotropy energy will be calculated numerically by assuming that $\lambda=1/4$ \cite{CSantos-JAP1}. In Fig. \ref{EnergiesDifStates} we present the magnetic energies for each magnetic state for different aspect ratio. In this figure  we can observe that for $p\leqslant0.7$ (thin torus), the TV  configuration minimizes the magnetic energy for any geometrical parameters of the torus shell. On the other hand, due to the increasing in the exchange energy in the TV configuration when $p$ increases, a VC configurations  becomes the minimum energy state when $p>0.75$ nm (Fig. \ref{EnergiesDifStates}b, c and d). In particular, since the exchange energy cost to keep the magnetic moments parallel to the surface in the region with negative curvature of the torus is very high, the VC state is highly stable and $\theta_c$ increases with $R$ (See Figs. \ref{EnergiesDifStates}c and d). In this context, the larger $R$, the larger the component of the magnetization pointing along the out-of-surface direction.  

\begin{figure}\begin{center}
\includegraphics[scale=0.3]{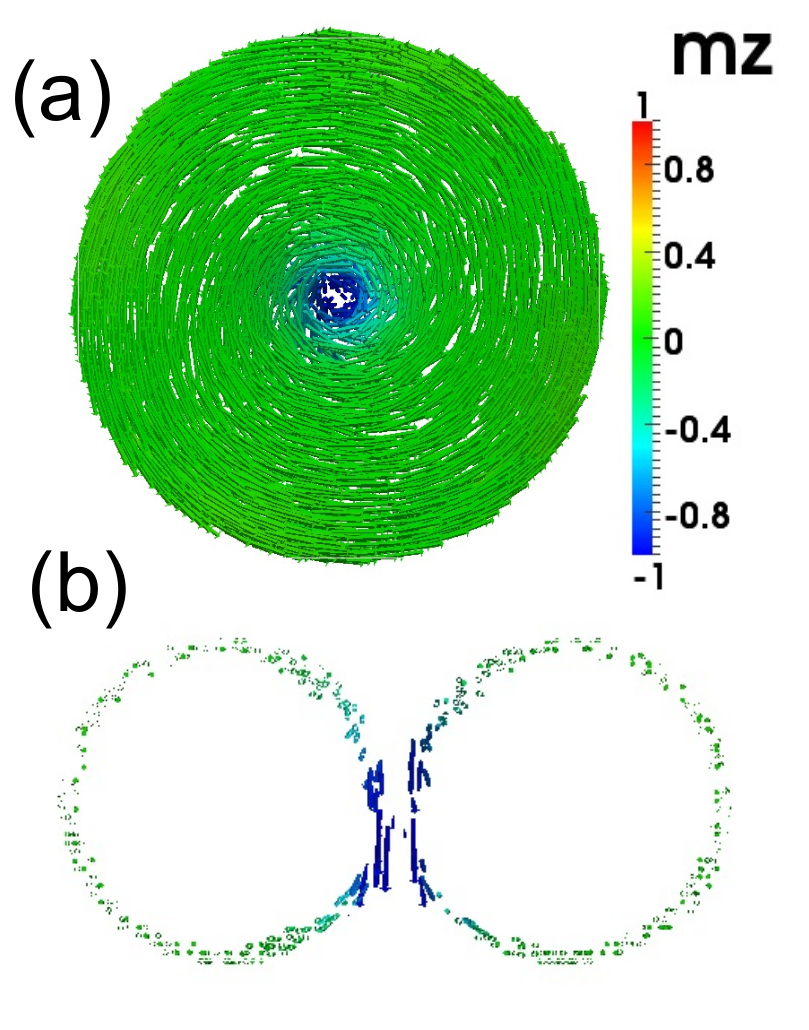}\caption{{Fig. a shows the magnetization field obtained from micromagnetic simulations for a nanotorus with $R=100$ nm and $r=99$ nm, seen from top. Fig. b) presents a cross section in the center of the nanotorus aiming to show the components of the magnetization pointing along the out-of-plane direction. }}\label{VCState}\end{center}
\end{figure}

{In order to corroborate our analytical results, we have performed micromagnetic simulations using OOMMF software\cite{oommf} in a Py nanotorus.  Simulated geometries consist in nanotori with internal and external radii ranging from 20 to 200 nm, and a thickness of the magnetic shell  of $0.2$ times the internal radii. In Fig. \ref{VCState}, we observed the VC states, and the core region in details for a nanotorus with $R=100$ nm and $r=99$ nm.}

\begin{figure}\begin{center}
\includegraphics[scale=0.5]{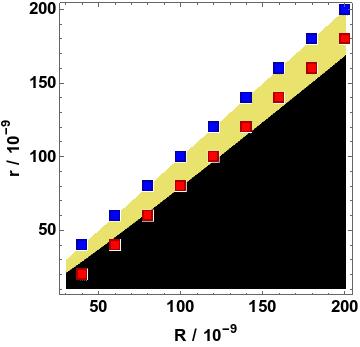}\caption{Phase diagram showing the groundstate configuration in function of geometrical parameters describing a magnetic nanotorus. Black and yellow regions represent respectively TV and VC configurations. Red points shows the transition line between the TV and VC states obtained from micromagnetic simulations. Blue points illustrate results from micromagnetic simulations for a the self-intersepting torus.}\label{StateDiagram}\end{center}
\end{figure}

The performed micromagnetic simulations and analytical calculations allow us to build a diagram state  showing the lowest energy state in toroidal magnetic shells as a function of the geometrical parameters $r$ and $R$. From the analysis of Fig. \ref{StateDiagram}, it can be noticed that there is a straight line dividing the regions in which TV and VC are stable. Analytical calculations are represented by continuous regions, while micromagnetic simulations are represented by dots. Black region identifies the toroidal geometrical parameters allowing a TV like magnetic state while yellow one represents the region in which VC state minimizes magnetic energy. The appearing of VC state as groudstate for $r/R\geqslant0.8$ is a result of the increasing in the exchange energy of VC state when $r\rightarrow R$. It is worth to say that despite we have obtained a very good agreement between analytical calculations and micromagnetic simulations, effects of the discretization and thickness of the simulated geometry yield the appearing of a small deviation in the inclination of the line defining the VC to TV state transition. In fact, the obtained line of transition between TV and VC states by means of micromagnetic simulations are represented by red dots. It is worth to say that in the obtained diagram state we do not consider geometrical parameters of toroidal nanoparticles allowing onion state as the minimum energy configuration \cite{CSantos-JAP1,Smiljan-JAP1}.

\section{Conclusions and Prospects}

By means of analytical calculations and micromagnetic simulations, we have analysed the static magnetic properties of a toroidal nanoshell. The comparison among the exchange energy of different in-surface states has shown that a magnetization configuration with the magnetic moments pointing along azimuthal direction (TV) is highly stable. The configuration with the magnetic moments pointing along polar-like direction (PV) is stable only when the aspect ratio $p=r/R\gtrsim0.87$. Nevertheless, the large dipolar energy of PV state makes this state unstable and TV state is the in-surface state that minimizes magnetic energy for any geometrial parameters of a magnetic toroidal shell.

We have also considered magnetic configurations in which the magnetic moments pointing along out-of-surface direction. Our results showed that toroidal magnetic nanoparticles can present a magnetic configuration resembling a vortex with a core (VC). It is shown that VC state appears as the magnetic groundstate for $p\gtrsim0.75$. It is also shown that the component of the magnetization pointing along out-of-surface direction increases whit $p$. The possibility of obtaining a VC state in toroidal nanoparticles is a very interesting result. In fact, the presence of such component can allow the manipulation of toroidal nanoparticles through an external magnetic field. Thus, such system could be used as drug delivery container \cite {JChem} and intracellular hyperthermia cancer therapy \cite{Cancer-1}. 

\section*{Acknowledgement}
In Brazil, this study was financed in part by the Coordenação de Aperfeiçoamento de Pessoal de Nível Superior - Brasil (CAPES) - Finance Code 001. The authors also thank CNPq (grant numbers 401132/2016-1 and 301015/2015-5) and FAPEMIG for financial support. In Chile, the authors acknowledge financial support from Fondecyt Grants 1160198, 1150072 and 11170858, Proyecto Basal USA 1555, and Financiamiento Basal para Centros Cient\'ificos y Tecnol\'ogicos de Excelencia (CEDENNA), under project FB0807.

\appendix 

\section{Geometrical properties of a torus}\label{Appendix1}

In this apendix we presented some aspects of toroidal geometry
that is used in the main tex.
Geometrically, a torus of genus-1 (one central
hole, descripted in Fig. \ref{Torus-Coordinates}) is a compact surface whose
Gaussian curvature smoothly varies from $1/r(R-r)$
to $+1/r(R+r)$ along its polar angle, so interpolating
between the pseudospherical and spherical curvatures
whenever $R>r$. From the topological point of view,
the simplest genus-1 torus is obtained by the topological
product of two circles, $T^1 = S^1 \wedge S^1$ . In general, a genus-n 
torus is obtained in a similar way, for example a
genus-2 is given by $T^2 = T^1 \wedge T^1$ . An ordinary torus is the
simplest two-dimensional closed manifold having genus-1, a single hole, like depicted in Fig. \ref{Torus-Coordinates}.
Depending on
the relative sizes of its internal, $r$, and external, $R$, radii,
three distinct standard types are obtained: ring ($R > r$), horn ($R = r$), or
self-intersecting spindle torus ($R < r$). The more common one is the ring
torus, resembling a donut, which is the type studied here, once the other ones, 
although mathematically interesting, seem to be rare in Nature. For describing
its geometry there are two coordinate systems frequently
used, the toroidal and the spherical-type coordinates, the last one relates the Cartesian to
the toroidal polar and azimuthal angles in a way similar to a spherical coordinates.

Our conventions are based in Ref.\cite{Gaididei-PRL}.
We considerer two dimensional surface ${\cal S}$ immersed in a three dimensional Euclidian 
space $R^3$ with the parametric representation in the form:
\be
\vec r=\vec r (q_1,\,q_2)
\ee
$q_\alpha$ are local curvilinear coordinates on the surface ${\cal S}$ and $\vec r$ is the three
dimensional position vector, in cartesian bases $\vec r= x\hat x + y\hat y + z\hat z$. Latin indices
$i,\,j,\,k\,,\ldots =1,\,2,\,3$ describe cartesian coordinates and greek indices 
$\alpha,\,\beta,\,\gamma,\,\ldots = 1,\,2$ curvilinear coordinates. The Einstein sum convention is
used, where repeated indices are summed. Any ordinary torus may be parametrized, for instance, by peripolar coordinates 
$(\theta,\,\varphi)$, so that:

\be
(R-\sqrt{x^2 + y^2})^2 + z^2 =r^2\,,
\ee
which are related to the Cartesian ones by:
\be
\label{parametricequation1}
\left\{\begin{array}{l}
x=\mathcal{Z}\cos\varphi\,,\\
y=\mathcal{Z}\sin\varphi\,,\\
z=r\cos\theta\,,
\end{array} \right.
\ee
where $R$ is the rotating (external), $r$ is the axial (internal) radii and the polar
and azimuthal angles are $\theta$ ($0 \leq \theta < 2\pi$) and $\varphi$ ($0 \leq \varphi < 2\pi$)  
see Fig. \ref{Torus-Coordinates}. The definition (\ref{parametricequation1}) generates the following geometrical 
properties of the surface ($(\alpha,\,\beta)=(\theta,\,\varphi)=(1,\,2)$).

Local normalized curvilinear basis:

\be
\hat e_\alpha=\frac{\vec g_\alpha}{|\vec g_\alpha|}=
\frac{\partial_\alpha\vec r}{|\partial_\alpha\vec r|}\,,\quad
\hat n =\hat e_1\times \hat e_2\,,
\ee

\ba
\hat g_\theta &=& \hat x r\cos\theta\cos\varphi + \hat y r\cos\theta \sin\varphi -
\hat z r\sin\theta\,,  \nonumber \\
\hat g_\varphi &=& -\hat x \mathcal{Z}\sin\varphi + \hat y \mathcal{Z}\cos\varphi\,,
\ea

\ba
\hat e_\theta &=& \hat x \cos\theta\cos\varphi + \hat y \cos\theta \sin\varphi -\hat z \sin\theta \,, 
\nonumber \\
\hat e_\varphi &=& -\hat x \sin\varphi + \hat y \cos\varphi \,,\\
\hat e_r &=& \hat x\sin\theta\cos\varphi +\hat y\sin\theta\sin\varphi +\hat z \cos\theta \,. \nonumber
\ea

Metric $g_{\alpha\beta}=\vec g_\alpha\cdot \vec g_\beta$:

\be\label{sphericalmetricL1}
{g_{\alpha\beta}}=
\left(\begin{array}{cc}
r^2  & 0   \\
0   & \mathcal{Z}^2      \\
\end{array} \right)\,,
\ee
being $\sqrt{g}=r\mathcal{Z}$. The spin connection 
$\Xi_\alpha=\hat\theta\cdot\partial_\alpha\hat\phi\sqrt{g_{\alpha\alpha}}$ 
also called first fundamental form:

\be
\Omega = -\frac{\cos\theta}{\mathcal{Z}}\hat{\varphi}\,.
\ee
The second fundamental form $b_{\alpha\beta}=\hat r\cdot\partial_\beta\vec g_\alpha$

\be\label{sphericalmetricL2}
{b_{\alpha\beta}}=
\left(\begin{array}{cc}
-r  & 0   \\
0   & - \sin\theta\mathcal{Z}      \\
\end{array} \right)\,,
\ee
and the Hessian matrix is

\be\label{sphericalmetricL3}
{h_{\alpha\beta}}=
-\left(\begin{array}{cc}
\frac1r  & 0   \\
0   & \frac{\sin\theta}{\mathcal{Z}}   \\
\end{array} \right)\,.
\ee
Gauss curvature ${\cal K}=det(h_{\alpha\beta})= \frac{\sin\theta}{r\mathcal{Z}}$ 
and the meam curvature ${\cal H}=Tr(h_{\alpha\beta})/2$. When the exchange energy density
is parametrized in the basis of a curved arbitrary thin magnetic shell, see eq. (\ref{EnFunc}), appear 
$\Gamma(\Phi)$, a vector depending on the Gauss and mean curvatures of the nanomagnet. Using the 
parametrization in Eq. (\ref{parametricequation1}) and the definitions of ${\Gamma(\Phi)}$ and $\Omega$,  
given in Ref. \cite{Gaididei-PRL}, Eqs. (4) and (5) can be promptly obtained.

\be
\Gamma(\Phi)=-\frac{\hat\theta}{r}\cos\Phi-\hat\varphi\frac{\sin\theta}{\mathcal{Z}}\sin\Phi\,.
\ee

Finally the differential operators (gradient, divergent and laplacian) in toroidal coordinates are:

\be
\mathbf\nabla f=\frac{\hat\theta}{r}\frac{\partial f}{\partial\theta}+
\frac{\hat\varphi}{\mathcal{Z}}\frac{\partial f}{\partial\varphi}
\ee

\ba
\mathbf\nabla\cdot \mathbf A &=&\frac{1}{r\mathcal{Z}}\frac{\partial}{\partial\theta}
[\mathcal{Z}A_\theta]+\frac{1}{\mathcal{Z}}\frac{\partial A_\varphi}{\partial\varphi}
\ea

\ba
\mathbf\nabla^2 f &=&\frac{1}{r^2\mathcal{Z}}\frac{\partial}{\partial\theta}\mathcal{Z}\frac{\partial f}{\partial\theta}
+\frac{1}{\mathcal{Z}^2}\frac{\partial^2 f}{\partial\varphi^2}
\ea



\begin{thebibliography}{99}


\bibitem{Nagaosa-Review}
N. Nagaosa, ant Y. Tokura, Nat. Nano. \textbf{8}, 899 (2013).

\bibitem{Gaididei-Review}
R. Streubel, P. Fischer, F. Kronast, V.P. Kravchuk, D.D. Sheka, Y. Gaididei, O.G. Schmidt, and D. Makarov, J. Phys. \textbf{D 49}, 363001
(2016).

\bibitem{CSantos-JAP1}
V.L. Carvalho-Santos, W.A. Moura-Melo, and A.R. Pereira, J. Appl. Phys. \textbf{108}, 094310 (2010).

\bibitem{Vagson-JAP-2015}
V.L. Carvalho-Santos, R.G. Elias, J.M. Fonseca, and D. Altbir, J. Appl. Phys. \textbf{117}, 17E518 (2015).

\bibitem{Kravchuk-PRB-2012} V.P. Kravchuk, D.D. Sheka, R. Streubel, D. Makarov, O.G. Schmidt, and Y. Gaididei, Phys. Rev. 
\textbf{B 85}, 144433 (2012).

\bibitem{Goll-PRB-2004} D. Goll, A. E. Berkowitz, and H. N. Bertram, Phys. Rev. B \textbf{70}, 184432 (2004).

\bibitem{New-sphere}
C. McKeever, F.Y. Ogrin and M.M. Aziz, J. Phys. D: Appl. Phys. \textbf{51}, 305003 (2018).

\bibitem{CSantos-PLA-2013} V. L. Carvalho-Santos, F. A. Apolonio, N. M. Oliveira-
Neto, Phys. Lett. A \textbf{377}, 1308 (2013).

\bibitem{CSantos-PLA-2012} V. L. Carvalho-Santos, and R. Dandoloff, Phys. Lett. A
\textbf{376}, 3551 (2012).

\bibitem{CSantos-PLA-2013-2}V. L. Carvalho-Santos, and R. Dandoloff, Braz. J. Phys.
\textbf{43}, 130 (2013).

\bibitem{Freitas-PLA-2005}W. A. Freitas, W. A. Moura-Melo, A. R. Pereira, Phys.
Lett. A \textbf{336}, 412 (2005).

\bibitem{Bose-Einstein} F. M. H. Crompvoets, H. L. Bethlem, R. T. Jongma, and G. Meijer, Nature (London) {\textbf 411}, 174 (2001); 
S. Gupta, K. W. Murch, K. L. Moore, T. P. Purdy, and D. M. Stamper-Kurn, Phys. Rev. Lett. {\textbf 95}, 143201 (2005).

\bibitem{nanotubo1} B. I. Dunlap, Phys. Rev. B {\textbf 46}, 1933 (1992); J. Liu, H. Dai, J. H. Hafner, D. T. Colbert, R. E. Smalley, S. T.Tans, and C. Dekker, Nature London {\textbf 385}, 780 (1997).

\bibitem{nanotubo2} S. Zhao, S. Zhang, M. Xia, E. Zhang, and X. Zuo, Phys. Lett. \textbf {A 331}, 238 (2004).

\bibitem{aplic-tecno} H. Watanabe, C. Manabe, T. Shigematsu, and M. Shimizu, Appl. Phys. Lett. \textbf{78}, 2928 (2001).


\bibitem{Jakson-EPJB-2016} J.M. Fonseca, V.L. Carvalho-Santos, A.R. Moura, W.A. Moura-Melo and A.R. Pereira  Eur. Phys. J. B \textbf{89}, 
153 (2016).

\bibitem{Mary-PRB-2007} A. Mary, D. M. Koller, A. Hobenau, J. R. Kren, A.
Boubelier, and A. Dereux, Phys. Rev. B \textbf{76}, 245422 (2007).

\bibitem{Mary-PRB-2005} A. Mary, A. Dereux, T. L. ferrell, Phys. Rev. B \textbf{72},
155426 (2005).

\bibitem{C-Santos08} V.L. Carvalho-Santos, A.R. Moura, W.A. Moura-Melo, and A.R. Pereira, Phys. Rev. B \textbf{77}, 134450 (2008).

\bibitem{Hazan-Mol-Torus}
M. Kowsar, and H. Sabzyan, Mol. Phys. \textbf{44}, 263 (2017).

\bibitem{Smiljan-JAP1}
S. Vojkovic, A.S. Nunez, D. Altbir, V.L. Carvalho-Santos, J. Appl. Phys. \textbf{120}, 033910 {2016}.

\bibitem{Gaididei-PRL}Y. Gaididei, V.P. Kravchuk, and D.D. Sheka, Phys. Rev. Lett. \textbf{112}, 257203 (2014).

\bibitem{Smiljan-JAP2} S. Vojkovic, V.L. Carvalho-Santos, J.M. Fonseca, A.S. Nunez,  J. Appl. Phys. \textbf{121}, 113906 (2017).

\bibitem{Kravchuck-ring}
V.P. Kravchuk, D.D. Sheka, and Y.B. Gaididei, J. Magn. Mag. Mat. \textbf{310}, 116 (2009).

\bibitem{Landeros-JAP}
P. Landeros, and \'A.S. N\'u\~nez, J. Appl. Phys. \textbf{108}, 033917 (2010).

\bibitem{Gaididei-JPA}
D.D. Sheka, V.P. Kravchuk, and Y. Gaididei, J. Phys. A: Math. Theor. \textbf{48}, 125202  (2015).



\bibitem{Benoit} J. Benoit and R. Dandoloff, Phys. Let. A \textbf{248}, 439 (1998).


\bibitem{Landeros-dot}
P. Landeros, J. Escrig, D. Altbir, D. Laroze, J. d’Albuquerque e Castro, and P. Vargas, Phys. Rev \textbf{B 71}, 094435 (2005)

\bibitem{oommf}
M. J. Donahue and D. G. Porter, National Institute of Standards and Technology Interagency Reports NISTIR No. 6376, (1999).

\bibitem{JChem}
S J. Son, J. Reichel, B. He, M. Schuchman, and S.B. Lee, J. Am. Chem. Soc. \textbf{127}, 7316 (2005).

\bibitem{Cancer-1}
C.S.B. Dias, T.D.M. Hanchuk, H. Wender, W.T. Shigeyosi, J. Kobarg, A.L. Rossi, M.N. Tanaka, M.B. Cardoso, and F. Garcia, Sci. Rep. \textbf{7}, 14843 (2017).

\end{thebibliography}
\end{document}